\documentclass[floats,floatfix,showpacs,amssymb,prd,twocolumn,superscriptaddress,nofootinbib,nolongbibliography,reprint]{revtex4-2}

\usepackage{amsmath}
\usepackage{amsfonts}
\usepackage{amssymb}
\usepackage{booktabs}
\usepackage[T1]{fontenc}
\usepackage{graphicx}
\usepackage[dvipsnames, usenames]{xcolor}
\definecolor{linkcolor}{rgb}{0.0,0.3,0.5}
\usepackage[unicode, colorlinks=true, linkcolor=linkcolor, citecolor=linkcolor, filecolor=linkcolor,urlcolor=linkcolor, pdfusetitle]{hyperref}
\usepackage{orcidlink}
\usepackage{siunitx}
\usepackage{slashed}
\usepackage[normalem]{ulem}

\newcommand{\milan}{Dipartimento di Fisica ``G. Occhialini'', 
Universit\`a degli Studi di Milano-Bicocca, Piazza della Scienza 3, 20126 Milano, Italy}
\newcommand{\infn}{INFN, Sezione di Milano-Bicocca, 
Piazza della Scienza 3, 20126 Milano, Italy}

\newcommand{\I}{\mathcal{I}}
\newcommand{\chiz}{\chi_{1\mathrm{z}}}
\newcommand{\hatchiz}{\hat{\chi}_{1\mathrm{z}}}

\newcommand{\Msun}{\mathrm{M}_{\odot}}
\newcommand{\Mtot}{M_{\mathrm{tot}}}
\newcommand{\Nobs}{N_{\mathrm{obs}}}

\begin{document}

\title{
    Exceptionality of exceptional gravitational-wave events
}

\author{Rodrigo Tenorio$\,$\orcidlink{0000-0002-3582-2587}}
\email{rodrigo.tenorio@unimib.it}
\affiliation{\milan}
\affiliation{\infn}

\author{Davide Gerosa$\,$\orcidlink{0000-0002-0933-3579}}
\email{davide.gerosa@unimib.it}
\affiliation{\milan}
\affiliation{\infn}

\begin{abstract}
In gravitational-wave astronomy, as in other scientific disciplines, ``exceptional'' sources attract considerable interest because they challenge our current understanding of the underlying (astro)physical processes. Crucially, “exceptionality” is defined only relative to the rest of the detected population. For instance, among all  gravitational-wave events detected so far, GW231123 is the binary black hole with the largest total mass, while GW241110 is the binary black hole with the most strongly misaligned spin relative to the orbital angular momentum.
\citeauthor{Mandel:2025qnh}~\cite{Mandel:2025qnh} argued that apparent ``exceptionality'' may reflect measurement error rather than an extreme true value, and suggested that the total mass of GW231123 may be significantly overestimated. Here we present a quantitative analysis that supports this conceptual point. We find that claims of “exceptionality” obtained under {population-agnostic} priors should be critically questioned whenever measurement uncertainties are comparable to the width of the underlying population. 
Specifically, we find that the total mass of GW231123 is unlikely to be meaningfully affected by this effect while the spin of GW241110 is far less likely to be anti-aligned than initially claimed: about 70\% of realizations that appear to yield an ``exceptionally anti-aligned'' spin are in fact consistent with either nonspinning or aligned configurations. 
\end{abstract}

\maketitle

\section{Introduction}
Gravitational-wave (GW) detection of black-hole (BH) binaries is now becoming routine. In this context, it is natural to single out some detected events as ``exceptional'' because of their informative properties, distinguishing them from the bulk of the population (and, in practice for researchers, motivating dedicated “single-event papers'').
Most often, such exceptional events correspond to extreme
values in some of the binary parameters (masses, spins, redshift) compared to the other events observed so far. 
For example, GW231123~\cite{LIGOScientific:2025rsn} is the most massive GW event detected at the time of writing, with a total mass of $\Mtot = 238^{+28}_{-49},\Msun$ (90\% credibility), and has sparked considerable interest because of its implications for, e.g., pair-instability processes in supernovae and hierarchical BH mergers. The previous record holder for the largest total mass from the preceding data-taking period, GW190521~\cite{LIGOScientific:2020iuh}, sparked similar interest at the time.
Events GW241011 and GW241110~\cite{LIGOScientific:2025brd} present the highest spin component aligned/anti-aligned with the binary orbital angular momentum,
respectively $\chiz = 0.66^{+0.08}_{-0.09}$ and $\chiz = -0.39^{+0.34}_{-0.37}$
(90\% credibility), with important consequences for the rate of binary BHs in dynamical formation channels.  

These measurements are obtained by conducting Bayesian inference on the
observed data $d$ using a waveform model parametrized by parameters $\theta$. Specifically,
the quantiles quoted above are summary statistics of the posterior distribution 
\mbox{$p(\theta|d, \I) \propto p(d | \theta, \I) p(\theta | \I)$},
where the likelihood $p(d | \theta, \I)$ encapsulates the generative process of the
data (including noise model, detector properties,  signal waveform model, etc.) and the prior
$p(\theta | \I)$ encapsulates our initial belief on the observed signal parameters.
Note that inference is conditioned on the broader context $\I$, which is typically omitted from the notation and collects any unstated but relevant assumptions included in the inference process~\cite{Jaynes:2003jaq}. 

\citeauthor{Mandel:2025qnh}~\cite{Mandel:2025qnh} recently highlighted that  exceptional events are so because their \emph{inferred} parameters are exceptional compared to a population. The emphasis on the word \emph{inferred} is important here: such inference (a.k.a. parameter estimation) is typically performed using {population-agnostic} assumptions. Let us denote this context by $\I_{\mathrm{ag}}$; GW practitioners commonly refer to this as ``the PE prior.'' Crucially, $\I_{\mathrm{ag}}$ {aims to simplify the analysis of individual events at the expense of population-level information (e.g. uniform prior on component masses).} We therefore find ourselves in a conundrum: an event is deemed ``exceptional'' relative to a population, yet the ``exceptionality'' itself ignores the population.
Such statements of exceptionality are therefore ill posed because the information required demands a different set of assumptions, $\I_{\mathrm{pop}}$ (through e.g. population-informed priors~\cite{Fishbach:2019ckx,Moore:2021xhn}).

In this paper, we examine the implications of this line of thought for some GW events that are currently quoted as ``exceptional,'' providing a quantitative analysis that complements Ref.~\cite{Mandel:2025qnh}. 

\section{Exceptional events and measurement errors}
Let us denote by $\theta$ a parameter of interest of a GW event (mass, spin, ...).
This parameter can be understood as a draw from a detected astrophysical population
$\theta \sim p(\theta | \mathrm{det}, \{d\})$, where $\{d\}$ denotes data from multiple events, and $\mathrm{det}$ refers to the fact that those
events passed a detectability threshold. 
For each event, all we know about the parameters $\theta$ is encapsulated in the posterior distribution $p(\theta|d, \I)$ from which one often extracts summary statistics $\hat{\theta}$ (such as mean, median, maximum a posteriori, etc.). In general, the summary statistic will be shifted with respect to the true value, i.e. {$\hat{\theta} \neq \theta$}

Given a collection of $j=1, \dots, \Nobs$ events 
(i.e. a catalog), an exceptional event, marked with $\star$, is defined as that which extremizes 
the summary statistic, i.e. 
\begin{equation}
\hat{\theta}^{\star}  = \max_j \,\hat{\theta}^j
\qquad {\rm or}  \qquad\hat{\theta}^{\star}  = \min_j \,\hat{\theta}^j
       \,.
        \label{eq:except}
\end{equation}
The crucial realization is that, depending on the typical spread of 
$p(\theta|\mathrm{det}, \{d\})$ and measurement errors, the ``exceptionality'' of an 
event may not only stem from the population, but also from the measurement errors.

\begin{figure}
	\includegraphics[width=\columnwidth]{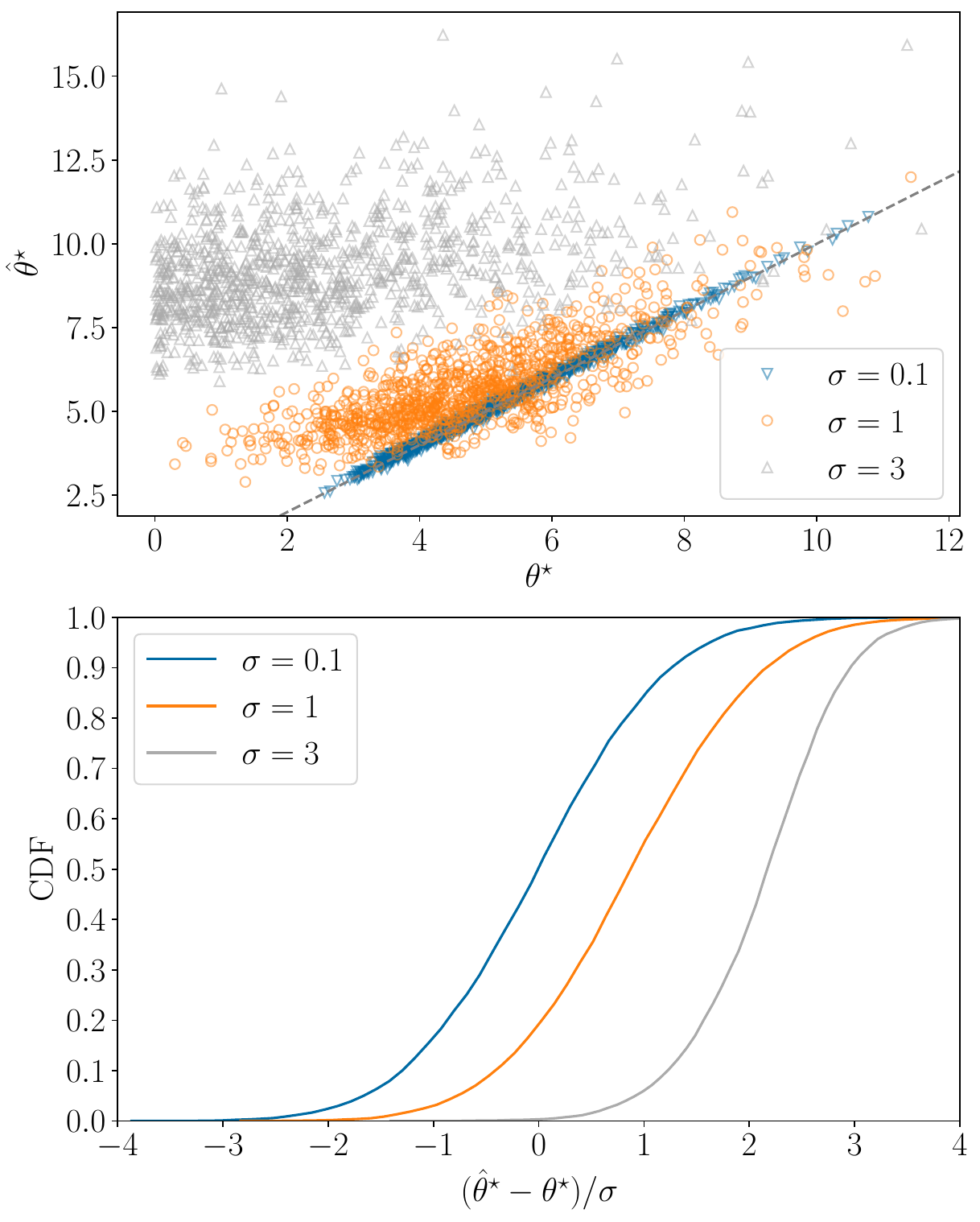}
    \caption{
        Distribution of exceptional events in a catalog of $N_{\mathrm{obs}}=100$
        using the toy population of Eq.~\eqref{eq:ppop} and
        Gaussian measurement errors. The parameter $\sigma$ controls the importance of measurement errors compared to the width of the population. The top panel compares the measured parameter $\hat\theta^\star$
	to the true value $\theta^\star$; the black dashed line corresponds to {$\hat{\theta}^\star=\theta^{\star}$}. The bottom panel shows the distribution of measurement errors.
    }
    \label{fig:toy_model}
\end{figure}

To illustrate this, suppose a single-parameter population following an exponential
distribution 
\begin{equation}
	p(\theta | \mathrm{det}, \{d\}) = 
	\begin{cases} 
		e^{-\theta} & {\rm if}\quad \theta \geq 0\,, \\
		0 & \mathrm{otherwise}\,, \\
	\end{cases}
    \label{eq:ppop}
\end{equation}
and measurement errors that are Gaussian with a standard deviation $\sigma$.
The population has unit expected value and unit standard deviation. 
The condition $\sigma > 1$ ($\sigma < 1$) corresponds to the case where the spread induced by measurement errors is larger (smaller) than the typical spread of the population. 
We generate 1000 realizations of a catalog with $N_{\mathrm{obs}}=100$ and collect
the highest mean $\hat{\theta}^{\star}$. 
Results from such a distribution of 
exceptional events are summarized in Fig.~\ref{fig:toy_model}.

The main result is that, for exceptional events, the summary statistics tends to significantly overestimate the
true parameter whenever the spread of measurement errors becomes compatible
with that of the population. For example, for $\sigma = 1$, we find that about 50\% of the measured
parameters overshoot the true value by more than $1\,\sigma$. 
For $\sigma = 3$, more than 90\% of 
the measured values are more than $1\,\sigma$ away, and more than 60\% are $2\,\sigma$ away. The latter corresponds to more than a factor of 2 of overestimation.

\section{GW231123}
\begin{figure}
    \includegraphics[width=\columnwidth]{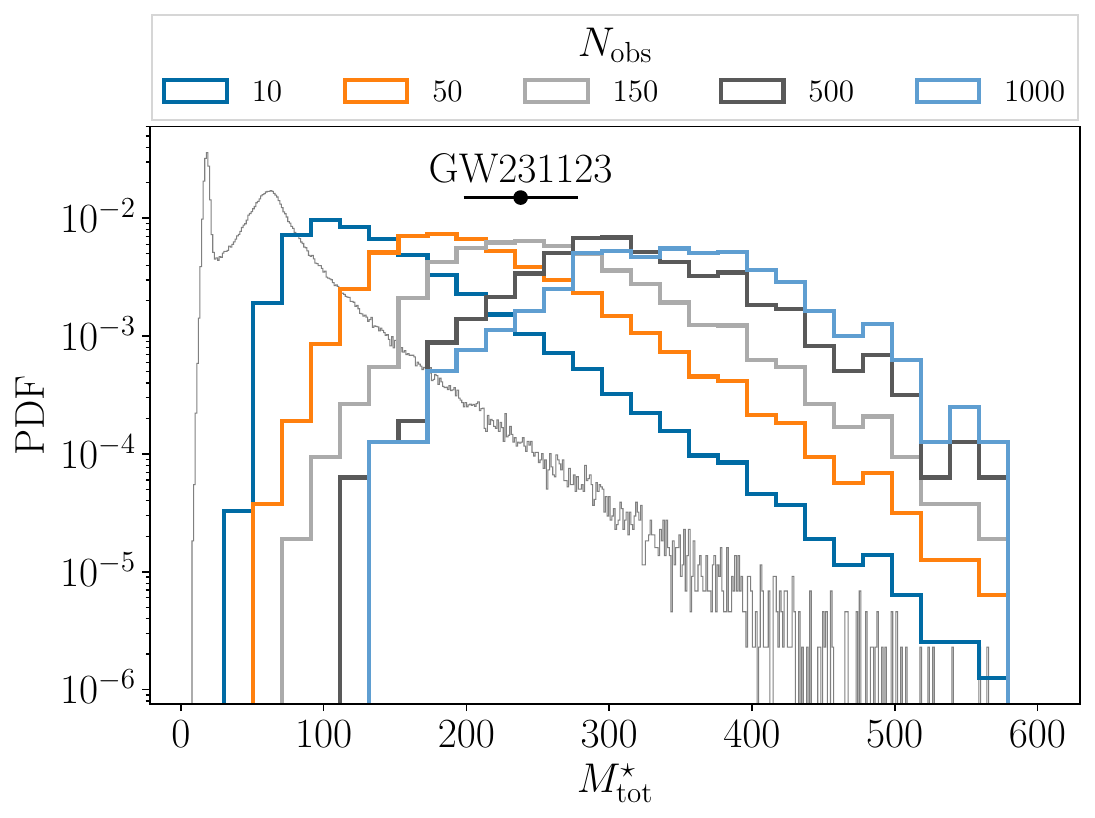}
    \caption{
        Exceptional events' total mass distribution using the GWTC-4.0 population as described in the main text. Catalogs of different sizes  $\Nobs$ are shown as colored histograms. The thin grey distribution
	shows the detected population $p(M_{\rm tot} | {\rm det}, \{d\})$. The scatter point indicates the total mass of GW231123 obtained under {population-agnostic} priors $\I_{\rm ag}$ from Ref.~\cite{LIGOScientific:2025rsn}.
    }
    \label{fig:Mtot_pop}
\end{figure}

We now apply this argument to the current set of GW events.
We simulate catalogs by sampling from the default binary BH population of GWTC-4.0~\cite{LIGOScientific:2025pvj} and imposing a detectability threshold corresponding to a network signal-to-noise (SNR) ratio of 8. We verified our results are not affected by this specific choice.
The SNR is computed using the power-spectral density and response of the
detectors in the event's observing run, which is sampled proportionally to the duration of 
the O1--O4 observing runs completed so far. As a cross-check,
we show in Fig.~\ref{fig:Mtot_pop} the events with largest total mass 
$M_{\rm tot}^\star =\max_j M_{\rm tot}^j$. For a catalog size comparable to GWTC-4.0
($\Nobs = 153$), the resulting distribution peaks at about the total mass of GW231123.

\begin{figure}
    \includegraphics[width=\columnwidth]{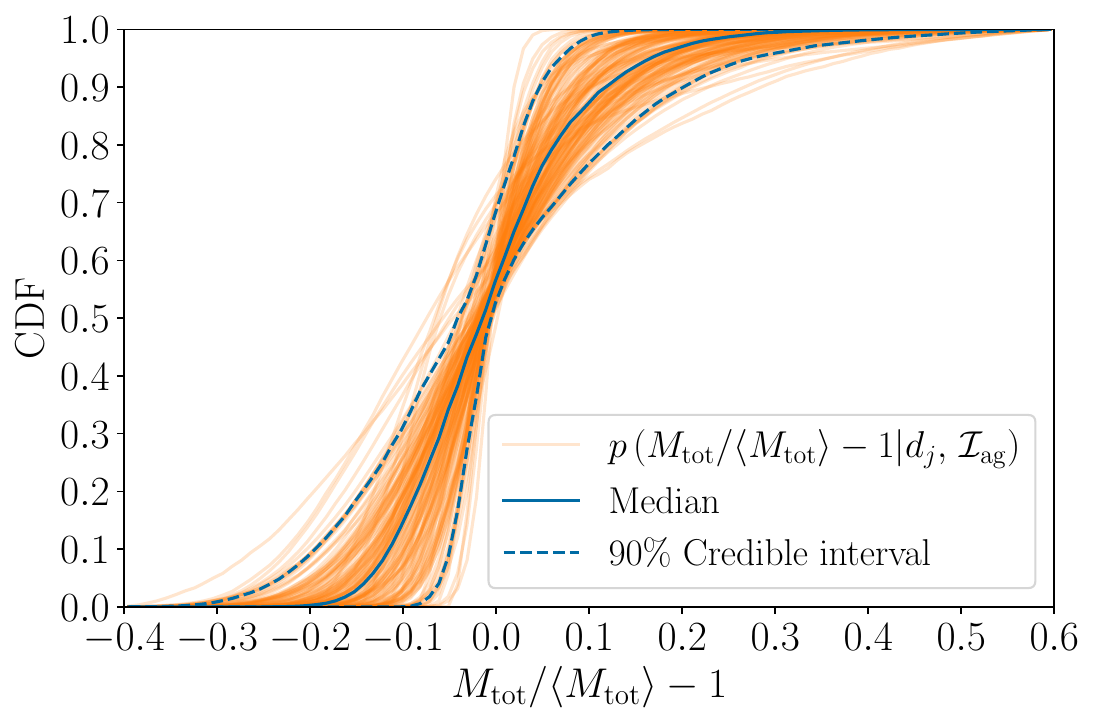}
    \caption{
	    {Distribution of relative errors in total mass with respect to the posterior mean for the 153
	    events considered from GWTC-4.0 using population-agnostic priors. 
	    The solid orange curves correspond to different events. The solid blue curve shows the median
	    of the distributions, while the dashed blue curves enclose the symmetric 90\% credible interval.}
    }
    \label{fig:Mtot_pop_rel}
\end{figure}

To construct the distribution of measurement errors, we standardize the 
{population-agnostic ($\mathcal{I}_{\mathrm{ag}}$)} posterior distribution 
on $M_{\mathrm{tot}}$ for all 153 binary BH events in GWTC-4.0 by subtracting 
and dividing by their posterior mean {$\langle M_{\mathrm{tot}} \rangle$,
as shown in Fig.~\ref{fig:Mtot_pop_rel}.
As a result, this procedure provides the distribution of relative deviations
$M_{\mathrm{tot}} / \langle M_{\mathrm{tot}} \rangle - 1$ for each event in the catalog. This quantity appears to be rather consistent: for all events, the relative deviation is bounded between about $-25\%$ 
and $+30\%$ at 90\% credibility.
This procedure is appropriate because shifts due to population reweighing are found to be typically contained within the posterior width (see, e.g., Fig.~2 of Ref.~\cite{LIGOScientific:2025slb} as well as Refs.~\cite{Moore:2021xhn,Mancarella:2025uat}). 
This paper exposes some of the issues associated with maximizing over events to identify exceptionality, for which all that is needed is a reasonable distribution of measurement errors.
}

\begin{figure}
    \includegraphics[width=\columnwidth]{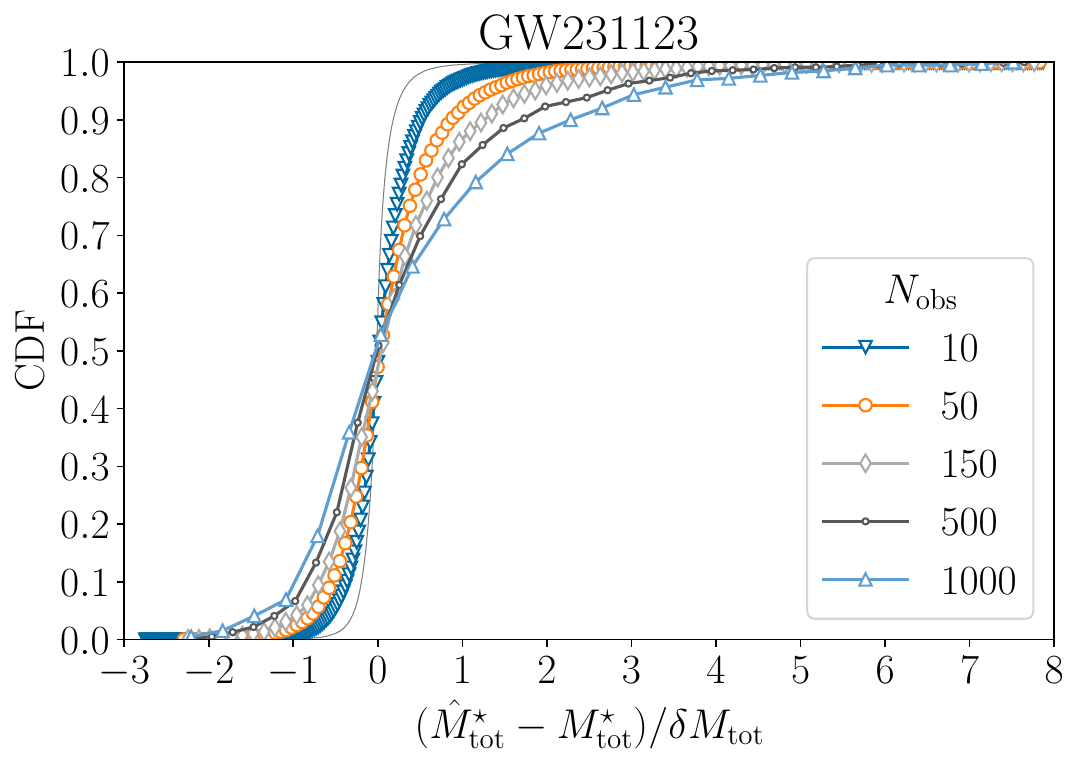}
    \caption{
        Deviation of the maximum measured total mass $\hat M_{\rm tot}^\star$ with respect to the true parameter $M_{\rm tot}^\star$ in units of $\delta M_{\rm tot}= 40 \Msun$. Results related to exceptional events from catalogs of sizes $N_{\rm obs}$ are shown with curves and markers. The thin grey curve indicates the corresponding distribution for the entire population $p(M_{\rm tot} | {\det}, \{d\})$ (i.e. without maximizing). This analysis illustrates the putative ``exceptionality'' of GW231123's total mass.
}
    \label{fig:Mtot_exceptional}
\end{figure}

Given a simulated event $M_{\mathrm{tot}}^{j}$, the corresponding measurement 
error is obtained by sampling a value from one of the 153 standardized posterior
distributions selected at random with equal probability.
The resulting distribution of exceptional events is shown in
Fig.~\ref{fig:Mtot_exceptional}, where we expressed {the} results in units of 
$\delta M_{\mathrm{tot}} = 40 \, \Msun$---a value indicative of  GW231123's uncertainty on the total mass.
{Note that the curves in Fig.~\ref{fig:Mtot_exceptional} are not symmetric: the summary statistic $\hat{M}_{\rm tot}^\star$ is more likely to overestimate than underestimate the true value $M_{\rm tot}^\star$ and, as expected, this asymmetry increases for larger catalogs. The effect in this case is overall rather mild because mass measurement errors are smaller 
than the size of the targeted population.}

One of the conclusions of~Ref.~\cite{Mandel:2025qnh} is that the true mass of 
GW231123 could be below $170\,\Msun$ as opposed to the value of 
$M_{\mathrm{tot}} = 238^{+28}_{-49}\,\Msun$ obtained using population-agnostic priors,
which corresponds to a shift of about $78\,\Msun = 1.95\,\delta M_{\mathrm{tot}}$.
We find that less than 5\% of our simulated catalogs with $\Nobs = 150$ support 
this conclusion.
While the typical spread of exceptional events is broader than that of the individual
events, the relatively small size of $\delta M_{\mathrm{tot}}$ with respect to the typical
range of $M_{\mathrm{tot}}$ suppresses the  ``exceptionality inflation.''

\section{GW241011 and GW241110}
\begin{figure}
    \includegraphics[width=\columnwidth]{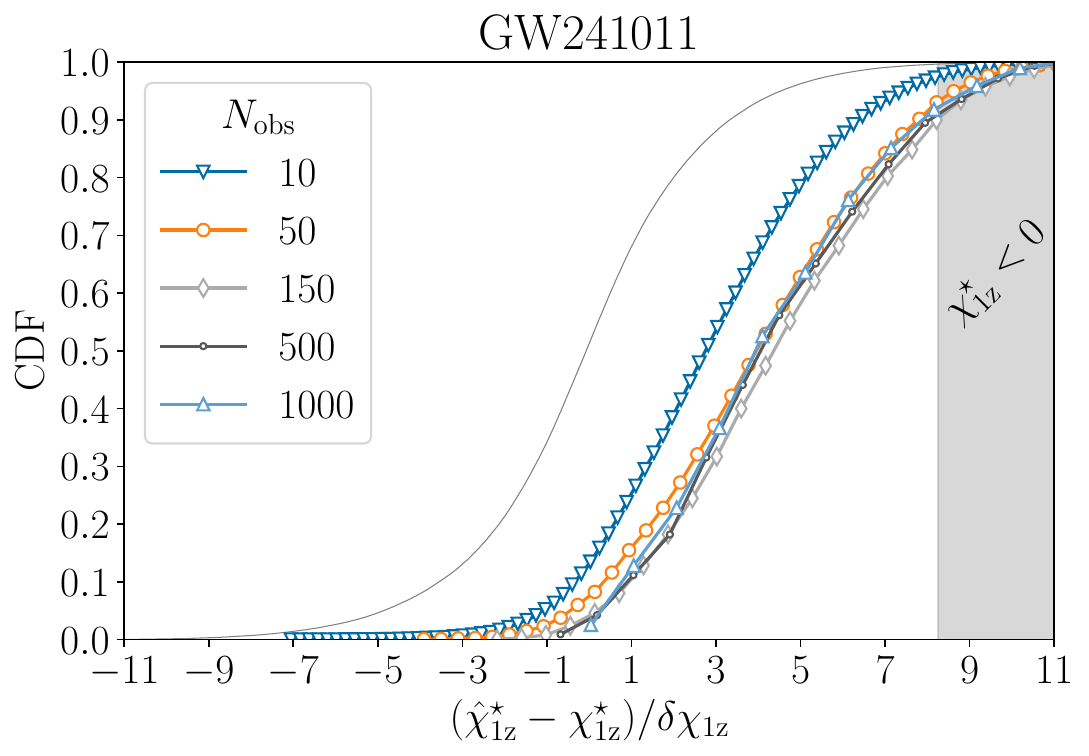}\\[5pt]
    \includegraphics[width=\columnwidth]{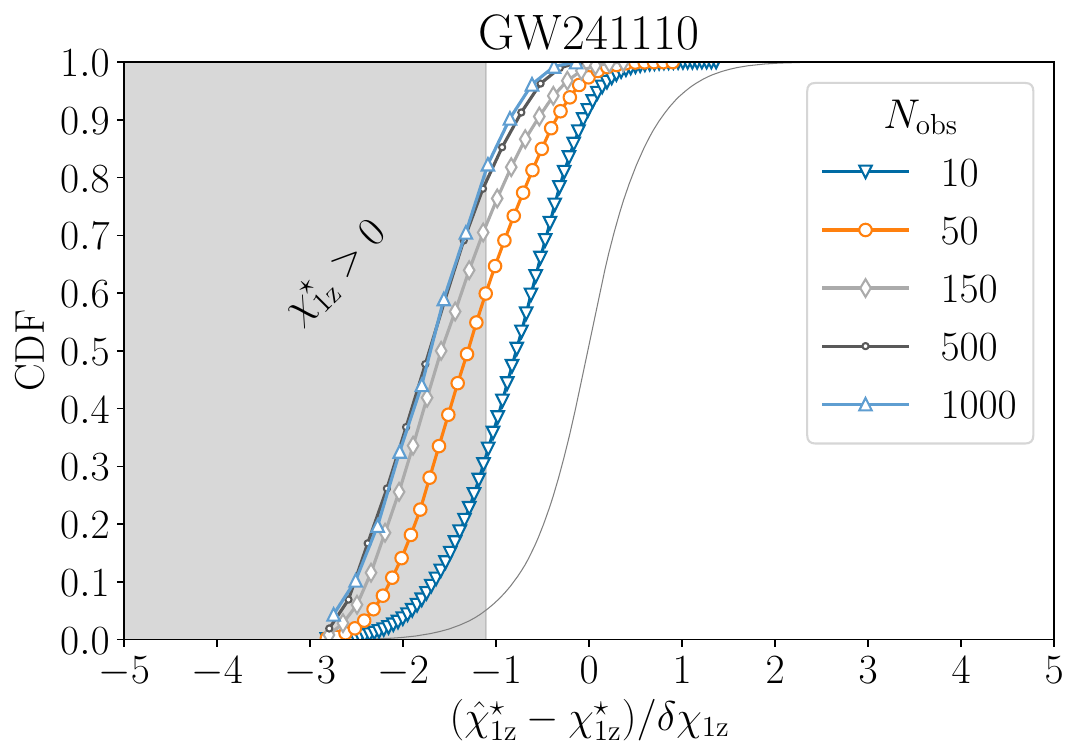}
    \caption{
               Deviation of the largest aligned/anti-aligned spin components  $\hat \chi_{1\mathrm{z}}^\star$ with respect to the true parameter $\chi_{1\mathrm{z}}^\star$. In the top (bottom) panel, we maximize (minimize) $\hat \chi_{1\mathrm{z}}$, which is indicative of GW241011 (GW241110) and use units of $\delta\chi_{1\mathrm{z}}=0.08$ ($\delta\chi_{1\mathrm{z}}=0.35$); see main text. Results related to exceptional events from catalogs of sizes $N_{\rm obs}$ are shown with curves and markers. The thin grey curves indicate the corresponding distribution for the entire population $p(\chi_{1\mathrm{z}} | {\det}, \{d\})$ (i.e. without extremizing). Grey areas correspond to cases where BH binaries with aligned spins are recovered with anti-aligned spins, and viceversa.
               }
    \label{fig:chi_exceptional}
\end{figure}

Similarly, we now consider the component $\chi_{1\mathrm{z}}$ of the primary BH spin aligned with the binary orbital angular momentum for the events GW241011 and GW241110~\cite{LIGOScientific:2025brd}. Since dimensionless BH spin components are bounded to the $(-1,1)$ interval, we expect measurement errors to strongly affect events with exceptional spin properties.
We simulate GW catalogs as described above. In this case, however,
posterior distributions are only shifted (and not re-scaled) by subtracting
the posterior mean, so that measurement errors correspond to (signed) absolute deviations, 
and measured  values are clipped to the $(-1,1)$ range after adding measurement errors.
Results are shown in Fig.~\ref{fig:chi_exceptional}.

GW241011 corresponds to the highest observed aligned spin
$\hatchiz^{\star} = \max_j \hatchiz^{j}  = 0.66$ with an uncertainty of about $\delta \chiz = 0.08$
(estimated from the reported 90\% credibility interval \cite{LIGOScientific:2025brd}, divided by 2).
About 90\% of our simulated catalogs
with $\Nobs \geq 50$ report a deviation of more than $1\,\delta\chiz$ with respect to 
the true value;  70\% of said catalogs return a deviation beyond $3\,\delta\chiz$.
{Only} 10\% of the catalogs return a deviation beyond $\approx 8.25\,\delta\chiz$, which
would imply GW241011 is non-spinning or even anti-aligned. 
{We conclude that 
GW241011's high-spin properties are unlikely to be driven by measurement errors. Note that this statement is conservative because the uncertainty associated to GW241011's $\chiz$ is significantly
smaller than that of a typical event in the catalog, which are used in the construction of our model.}

GW241110 corresponds to the highest observed anti-aligned spin
{$\hatchiz^{\star} =\min_{j} \hatchiz^{j}= -0.39$}.
The uncertainty in this case is about $\delta \chiz = 0.35$~\cite{LIGOScientific:2025brd}.
For $N_{\mathrm{obs}} = 150$, about 80\% of the exceptional events are more than $1\,\delta\chi_{1\mathrm{z}}$ away from their true parameter values, and 70\% of the exceptional events are compatible with BH spins that are preferentially aligned with the binary orbital angular momentum.

\section{Discussion}
BHs with masses of $\sim 30\,\Msun$ were considered exceptional ten years ago, following the detection of GW150914 \cite{LIGOScientific:2016aoc}, because such values were much higher than those inferred from electromagnetic observations at the time.  Today, BHs of $\sim 30\,\Msun$ are considered vanilla \cite{LIGOScientific:2025pvj}. This evolution is bound to continue. As the catalog of detected events grows, the distribution of what is deemed “exceptional” inevitably shifts toward more extreme values.
The same argument also applies to measurement errors: as we collect more and more data, it becomes increasingly likely to observe an extreme measurement error. This follows fundamentally from the use of population-agnostic priors to identify exceptional events, which is incompatible with the fact that events are deemed exceptional only with respect to a population~\cite{Mandel:2025qnh, Fishbach:2019ckx, Moore:2021xhn}.

In this work, we have presented a simplified model of the interplay between extreme events in a population and extreme measurement errors. We conclude that, for current catalog sizes $N_{\mathrm{obs}} \lesssim 10^3$, measurement errors drive apparent exceptional events only when their magnitude is comparable to the span of the population. This is the case for spins with uncertainties of $\delta\chi_{1\mathrm{z}} \approx 0.35$, such as GW241110, but it is unlikely to be a concern for GW241011 or for the most massive event observed to date, GW231123.

On the latter, our results are at odds with those of Ref.~\cite{Mandel:2025qnh}, which suggest that the total mass of GW231123 is likely to be significantly overestimated due to the use of inconsistent priors. While we agree with their overall argument, our quantitative analysis indicates that the typical uncertainty on the total mass renders this effect smaller than initially suggested. Specifically, we find that the 90\% credible interval reported under $\I_{\mathrm{ag}}$~\cite{LIGOScientific:2025rsn} is still likely to contain the true mass, despite the use of information that is inconsistent with respect to the population.

On the other hand, our analysis suggests that GW241110, the most confidently anti-aligned spin measured in with GWs so far under  $\I_{\mathrm{ag}}$
($\chi_{1\mathrm{z}}<0$ at 97.7\% credibility~\cite{LIGOScientific:2025brd}), may actually not be so.
Specifically, 70\% of our simulated catalogs return ``exceptional'' anti-aligned events 
which are in fact non-spinning/aligned.  This has major 
astrophysical implications, as spin is often quoted as a key smoking gun of  BH binary formation channels. 

Echoing the key message of Ref.~\cite{Mandel:2025qnh}, our results suggest caution when interpreting astrophysically interesting events in a population-agnostic manner, as the resulting conclusions may not be robust. 
In addition to Ref.~\cite{Mandel:2025qnh}, we stress this caveat is important whenever measurement errors are comparable in size to the width of the underlying source population.

\section{Acknowledgments}
This work was partly motivated by discussions related to a bet signed by Ilya Mandel, Thomas Callister, Daniel Holz, Davide Gerosa, and Jonathan Gair at the \emph{``Gravitational-wave snowballs, populations, and models''} workshop held in Sexten, Italy, in January 2025.  
We thank
Ssohrab Borhanian,
Federico De Santi,
Giulia Fumagalli,
Ana Lorenzo-Medina,
Matthew Mould,
and Alexandre Toubiana
for discussions.
R.T. and D.G. are supported by 
ERC Starting Grant No.~945155--GWmining, 
Cariplo Foundation Grant No.~2021-0555, 
MUR PRIN Grant No.~2022-Z9X4XS, 
Italian-French University (UIF/UFI) Grant No.~2025-C3-386,
MUR Grant ``Progetto Dipartimenti di Eccellenza 2023-2027'' (BiCoQ),
and the ICSC National Research Centre funded by NextGenerationEU. 
D.G. is supported by MSCA Fellowship No.~101149270--ProtoBH and MUR Young Researchers Grant No. SOE2024-0000125.
Computational work was performed at CINECA with allocations through INFN, 
the University of Milano-Bicocca, and ISCRA Type-C project HP10C29O69.

\bibliography{references}

\end{document}